# Bayesian spatial transformation models with applications in neuroimaging data

**Michelle F. Miranda**[1,*], **Hongtu Zhu**[2,**], and **Joseph G. Ibrahim**[2,***]

[1]Department of Statistics and Operations Research, University of North Carolina at Chapel Hill Chapel Hill, North Carolina, U.S.A.

[2]Department of Biostatistics, University of North Carolina at Chapel Hill Chapel Hill, North Carolina, U.S.A.

## Summary

The aim of this paper is to develop a class of spatial transformation models (STM) to spatially model the varying association between imaging measures in a three-dimensional (3D) volume (or 2D surface) and a set of covariates. Our STMs include a varying Box-Cox transformation model for dealing with the issue of non-Gaussian distributed imaging data and a Gaussian Markov Random Field model for incorporating spatial smoothness of the imaging data. Posterior computation proceeds via an efficient Markov chain Monte Carlo algorithm. Simulations and real data analysis demonstrate that the STM significantly outperforms the voxel-wise linear model with Gaussian noise in recovering meaningful geometric patterns. Our STM is able to reveal important brain regions with morphological changes in children with attention deficit hyperactivity disorder.

## Keywords

## 1. Introduction

The emergence of various imaging techniques has enabled scientists to acquire high-dimensional imaging data to closely explore the function and structure of the human body in various imaging studies. Several common imaging techniques include magnetic resonance image (MRI), functional MRI, diffusion tensor image (DTI), positron emission tomography (PET), and electroencephalography (EEG), among many others. These imaging studies, such as the Alzheimer's Disease Neuroimaging Initiative (ADNI), are essential to understanding the neural development of neuropsychiatric and neurodegenerative disorders, the normal brain and the interactive effects of environmental and genetic factors on brain structure and function, among others. A common feature of all these imaging studies is that they are generating many very high dimensional and complex data sets.

*fmiranda@email.unc.edu
**hzhu@bios.unc.edu
***ibrahim@bios.unc.edu

There is a great interest in developing voxel-wise methods to characterize varying associations between high-dimensional imaging data and low-dimensional covariates (Friston, 2007; Lindquist, 2008; Lazar, 2008; Li et al., 2011). These methods usually fit a general linear model to the imaging data from all subjects at each voxel as responses and clinical variables, such as age and gender, as predictors. Subsequently, a statistical parametric map of test statistics or *p*-values across all voxels (Lazar, 2008; Worsley et al., 2004) is generated. Several popular neuroimaging software platforms, such as statistical parametric mapping (SPM) (www.fil.ion.ucl.ac.uk/spm/) and FMRIB Software Library (FSL) (www.fmrib.ox.ac.uk/fsl/), include these voxel-wise methods as their key statistical tools.

These voxel-wise methods have several major limitations. Firstly, the general linear model used in the neuroimaging literature usually assumes that the imaging data conform to a Gaussian distribution with homogeneous variance (Ashburner and Friston, 2000; Wager et al., 2005; Worsley et al., 2004; Zhu et al., 2009). This distributional assumption is important for the valid calculation of *p*-values in conventional tests (e.g., F test) that assess the statistical significance of parameter estimates. Moreover, methods of random field theory (RFT) that account for multiple statistical comparisons depend strongly on the parametric assumptions, as well as several additional assumptions (e.g., smoothness of autocorrelation function).

Second, the Gaussian assumption is known to be flawed in many imaging datasets (Ashburner and Friston, 2000; Salmond et al., 2002; Luo and Nichols, 2003; Zhu et al., 2009). It is common to use a Gaussian kernel with the full-width-half-max (FWHM) in the range of 8–16 mm to account for registration errors, to make the data normally distributed and to integrate imaging signals from a region, rather than from a single voxel. However, recent research has shown that varying filter sizes in the smoothing methods can result in different statistical conclusions about the activated and deactivated regions, and spatial smoothing biases the localization of brain activity. Thus, it can result in misleading scientific inferences (Jones et al., 2005; Sacchet and Knutson, 2012).

Third, as pointed out in the literature (Li et al., 2011; Yue et al., 2010), the voxel-wise methods treat all voxels as independent units, and thus they ignore important spatial smoothness observed in imaging data. Several promising methods have been proposed to accommodate the varying amount of smoothness across the imaging space by using function-on-scalar regression in the functional data analysis framework (Zhu et al., 2012; Ramsay and Silverman, 2005; Staicu et al., 2010), adaptive smoothing methods within a frequentist framework (Polzehl and Spokoiny, 2006; Li et al., 2011), and spatial priors within the Bayesian framework (Gossl et al., 2001; Penny et al., 2005; Bowman et al., 2008; Smith and Fahrmeir, 2007). However, according to the best of our knowledge, none of them address the two issues including spatial smoothness and the Gaussian assumption simultaneously.

The aim of this paper is to develop a class of spatial transformation models (STMs) to simultaneously address the issues discussed above for the spatial analysis of neuroimaging data given a set of covariates. Our spatial transformation model is a hierarchical Bayesian model. First, we use a Box-Cox transformation model on the response variable assuming an unknown transformation parameter in order to satisfy the normality assumption in the imaging data, and then develop a regression model to characterize the association between the imaging data and the covariates. Second, we use a Gaussian Markov random field (GMRF) prior to capture the spatial correlation and spatial smoothness among the regression coefficients in the neighboring voxels. We develop an efficient Markov chain Monte Carlo (MCMC) algorithm to draw random samples from the desired posterior distribution. Our simulations and real data analysis demonstrate that STM significantly outperforms the standard voxel-wise model in recovering meaningful regions.

The rest of paper is organized as follows. In Section 2, we introduce the STM and its associated prior distributions and Bayesian estimation procedure. In Section 3, we compare STM with the standard voxel-wise method using simulated data. In Section 4, we apply STM to a real imaging dataset on attention deficit hyperactivity disorder (ADHD). Finally, in Section 5, we present some concluding remarks.

## 2. Model

### 2.1 Model Description

Consider imaging measurements in a common space, which can be either a 3D volume or a 2D surface, and a set of clinical variables (e.g., age, gender, and height) from $n$ subjects. Let $\mathcal{D}$ and $d$, respectively, represent the set of grid points in the common space and the center of a voxel in $\mathcal{D}$ and $N_D$ equals the number of voxels in $\mathcal{D}$ Without loss of generality, $\mathcal{D}$ is assumed to be a compact set in $\Re^3$. For the $i$-th subject, we observe a univariate imaging measure $y_i(d)$ at $d \in \mathcal{D}$ and an $N_D \times 1$ vector of imaging measures, denoted by $\mathbf{Y}_{i,\mathcal{D}} = \{y_i(d) : d \in \mathcal{D}\}$. For simplicity, we consider a 3D volume throughout the paper.

We propose a class of spatial transformation models consisting of two major components as follows: a transformation model and a Gaussian Markov random field model. The transformation model is developed to characterize the association between the imaging measures and the covariates at any $d \in \mathcal{D}$ and to achieve normality. Since most imaging measures are positive, we consider the well-known Box-Cox shifted power transformation (Box and Cox, 1964) throughout. Extensions to other parametric transformations are trivial (Sakia, 1992). Let $y_i(d)^{(\lambda)}$ be the Box-Cox transformation of $y_i(d)$ given by

$$y_i(d)^{(\lambda)} = \begin{cases} \{(y_i(d) + c_0)^{\lambda} - 1\}/\lambda, & \text{if } \lambda \neq 0, \\ \log(y_i(d) + c_0), & \text{if } \lambda = 0, \end{cases}$$

where $c_0$ is prefixed and chosen such that $\inf_{i,d}(y_i(d)) > -c_0$. Our Box-Cox transformation model is given by

$$y_i^{(\lambda_d)}(d) = \mathbf{x}_i^T \beta(d) + \varepsilon_i(d) \text{ for } d \in \mathcal{D}, \quad (1)$$

where $\beta(d) = (\beta_1(d), \ldots, \beta_p(d))^T$ is a $p \times 1$ vector of regression coefficients of interest, $\mathbf{x}_i$ is a $p \times 1$ vector of observed covariates for subject $i$, and $\varepsilon(d) = (\varepsilon_1(d), \ldots, \varepsilon_n(d))^T$ is an $n \times 1$ vector of measurement errors and follows a $N_n(0, \sigma^2(d)I_n)$ distribution, in which $I_n$ is an $n \times n$ identity matrix.

The Gaussian Markov random field (GMRF) model is proposed to capture the spatial smoothness and correlation for each component of $\{\beta(d) : d \in \mathcal{D}\}$ across all voxels. Moreover, by imposing GMRF for $\{\beta(d) : d \in \mathcal{D}\}$, we have implicitly modeled the spatial correlations among imaging measures across voxels. For $k = 1, \ldots, p$, $\beta_{(k)} = \{\beta_k(d) : d \in \mathcal{D}\}$ is defined to be the coefficient set associated with the $k$-th covariate across all voxels. In practice, it is very natural to assume that different $\beta_{(k)}$ images may have different patterns, since different covariates play different roles in characterizing their association with the imaging data. Specifically, we consider a GMRF for each $\beta_{(k)}$ by assuming that

$$\beta_{(k)} \sim N(\mathbf{0}, \nu_k^{-1}(I_{N_D} + \phi_k H_k)^{-1}),$$

where $\nu_k > 0$ and $\varphi_k > 0$ are, respectively, scale and spatial parameters. When $\varphi_k = 0$, the elements of $\boldsymbol{\beta}_{(k)}$ are independent, whereas when the value of $\varphi_k$ is large, the model approaches an intrinsic autoregressive model (Ferreira and De Oliveira, 2007; Rue and Held, 2005). The known matrix $H_k = \{h_k(d, d')\}$ is an $N_D \times N_D$ matrix allowing the modeling of different patterns of spatial correlation and smoothness. Let $N(d)$ be a set of neighboring voxels of voxel $d$ in a given neighborhood system. Using the properties of GMRF (Rue and Held, 2005), the full conditional distribution of $\beta_k(d)$ can be written as

$$\beta_k(d)|\beta_{(k),[d]}, \nu_k, \phi_k \sim N\left(\frac{\phi_k \sum_{d' \in N(d)} h_k(d, d')\beta_k(d')}{1 + \phi_k h_k(d, d)}, \frac{1}{\nu_k[1 + \phi_k h_k(d, d)]}\right), \quad (2)$$

where $\boldsymbol{\beta}_{k,[d]}$ contains all $\beta_k(d')$ for all $d' \in \mathscr{D}$ except $d$. The conditional mean of $\beta_{(k)}(d)$ is a weighted average of the $\beta_k(d')$ values in the neighboring voxels of $d$. As the number of neighboring voxels increases, the conditional variance decreases (Ferreira and De Oliveira, 2007).

A challenging issue is how to specify $H_k = \{h_k(d, d')\}$ for each $\boldsymbol{\beta}_{(k)}$ in order to explicitly incorporate the spatial correlation and smoothness among neighboring voxels. We set

$$h_k(d, d') = \begin{cases} \sum_{d' \in N(d)} \omega_k(d, d')^2, & \text{for } d = d', \\ -\omega_k(d, d')^2 \mathbf{1}(d' \in N(d)), & \text{for } d \neq d', \end{cases}$$

where $\omega_k(d, d')$ are some pre-calculated weights and $\mathbf{1}(A)$ is the indicator function of a set A. For every $\varphi_k \geq 0$, $(\mathbf{I}_{N_D} + \varphi_k H_k)^{-1}$ is diagonally dominant and thus positive definite. For computational efficiency, we choose a relatively small neighborhood for each voxel $d$ by defining $N(d) = \{d' : \|d - d'\|_2 \leq r_0\}$, where $r_0$ is a positive scalar and $\|\cdot\|^2$ denotes the Euclidean distance. There are several ways of choosing the weights $\omega_k(d, d')$ for any $d, d' \in \mathscr{D}$ Ideally, $\omega(d, d')$ should contain some similarity information, such as spatial distance and imaging similarity, between voxels $d$ and $d'$. The simplest example of $\omega_k(d, d')$ is $\omega_k(d, d') = K(\|d - d'\|_2)$, where $K(u) = \exp(-0.5u^2)\mathbf{1}(u \leq r_0)$. Other choices of $\omega_k(d, d')$ are definitely possible. For instance, one may borrow information learned from different imaging data and historic information in order to construct the similarity between $d$ and $d'$.

## 2.2 Priors

We first consider the priors for the remaining parameters in the first level of model (1). Let $\tau_d = (\sigma^2(d))^{-1}$ and $U(-a, b)$ denote the uniform distribution on the interval $(-a, b)$. We Specifically assume that for $d \in \mathscr{D}$

$$\tau_d \sim \text{Gamma}(\delta_0/2, \gamma_0/2) \text{ and } \lambda_d \sim \text{U}(-a, b).$$

For the second level parameter $\nu = (\nu_1, \ldots, \nu_p)$, we assume for $k = 1, \ldots, p$

$$\nu_k \sim \text{Gamma}(n_\nu/2, n_\nu s_\nu^2/2),$$

where $n_\nu$ and $s_\nu^2$ are hyperparameters. The choice of Gamma priors for the precision parameters is common in the literature since it maintains conjugacy (Chen et al., 2000).

Other choices are $\pi(\tau_d) = \tau_d^{-1}\mathbf{1}(\tau_d \geq 0)$ and $\pi(\nu_k) = \nu_k^{-1}\mathbf{1}(\nu_k \geq 0)$, which are improper but in both cases lead to a proper posterior distribution. The uniform prior for the transformation $\lambda_d$ was first introduced by Box and Cox (1964) and later adopted by several authors (Sweeting, 1984; Gottardo and Raftery, 2006).

### 2.3 Posterior Computation

An efficient Gibbs sampler is proposed to generate a sequence of random observations from the joint posterior distribution $p(\boldsymbol{\beta}, \lambda, \tau_\sigma, \nu | Y)$. The Gibbs sampler essentially involves sampling from a series of conditional distributions while each of the modeling components is updated in turn. Although the order of the parameter update does not affect convergence, updating the higher level parameters first can result in an improvement of the speed of convergence. Details pertaining to each step are presented below.

i.    Update each component of $\nu = (\nu_1, \ldots, \nu_p)$ from its full conditional distribution,

$$p(\nu_k | -) \sim \mathrm{Gamma}(0.5 n_{\nu_k}^*, 0.5 n_{\nu_k}^* s_{\nu_k}^{*2}),$$

where $\nu_k^* = N_D + n_\nu$ and $n_\nu^* s_\nu^{*2} = n_\nu s_\nu^2 + \sum_{j=1}^{N_D} \beta_k(j)^2 + \phi_k \beta_k^T H_k \beta_k$.

ii.    Update $\beta_{(k)}(d)$, $k = 1, \ldots, p$, for each voxel $d \in \mathscr{D}$ from its full conditional distribution,

$$p(\beta_{(k)}(d) | -) \sim \mathrm{N}(\mu_\beta(d), \sigma_\beta^2(d)),$$

where $\sigma_\beta^2(d) = (\tau_d \sum_i x_{ik}^2 + \theta_k(d))^{-1}$ and

$$\mu_\beta(d) = \sigma_\beta^2(d)\{\tau_d \sum_i [y_i^{(\lambda d)}(d) - \sum_{l \neq k} x_{il}\beta_l^{(m)}(d)]x_{ik} + \theta_k(d)m_{(k)}(d)\}.$$

Moreover, $\boldsymbol{\beta}^{(m)}(d)$ is the estimated value of $\boldsymbol{\beta}(d)$ obtained in the previous iteration of the Gibbs sampler and $\theta_k(d)$ and $m_{(k)}(d)$ are, respectively, the inverse of the variance and the mean of the Gaussian distribution in (2).

iii.    Update $\tau_\sigma(d)$ for each voxel $d \in \mathscr{D}$ from its full conditional distribution

$$p(\tau_\sigma(d) | -) \sim \mathrm{Gamma}\left(\frac{1}{2}(n + \delta_0), \frac{1}{2}\sum_i (y_i^{(\lambda_d)}(d) - \mathbf{x_i^T}\beta(d))^2 + \gamma_0\right).$$

iv.    Update $\lambda_d$ for each voxel $d \in \mathscr{D}$ from its full conditional distribution

$$p(\lambda_d | -) \sim \prod_{d \in D} \exp\left\{-\frac{\tau_d}{2}\sum_i (y_i^{(\lambda_d)}(d) - \mathbf{x_i^T}\beta(d))^2\right\} \times \prod_{i=1}^n y_i^{\lambda_d - 1}(d) \times (b + a)^{-1}.$$

The full conditional distribution of $\lambda_d$ does not have a closed form, but sampling methods such as the Slice Sampler (Neal, 2003) or the Adaptive Rejection Metropolis Sampling (ARMS) (Gilks et al., 1995) can be used for such a purpose. The Metropolis-Hastings (MH) algorithm (Hastings, 1970) is also a very useful and easy algorithm for sampling $\lambda_d$. The MH algorithm proceeds as follows:

**a.** Generate $\lambda_d^{prop}$ from $N(\lambda_d^{(t-1)}, \delta_\lambda)$ where $\delta_\lambda > 0$ is a tuning parameter.

**b.** Generate $V$ from $U(0, 1)$.

**c.** Let
$$\alpha = \min\left\{1, \frac{L(\lambda_d^{prop}, \beta_{(d), \tau_\gamma(d)}; \mathbf{Y}(\mathbf{d}))\pi(\lambda_d^{prop})}{L(\lambda_d^{(t-1)}, \beta_{(d), \tau_\gamma(d)}; \mathbf{y}(\mathbf{d}))\pi(\lambda_d^{(t-1)})}\right\}.$$

If $V \leq \alpha$, then set $\lambda_d^{(t)} = \lambda_d^{prop}$. Otherwise, set $\lambda_d^{(t)} = \lambda_d^{(t-1)}$.

## 3. Simulation Study

We carried out a simulation study to examine the finite-sample performance of the STM in establishing an association between the imaging data and a set of covariates. The goals of this simulation study are

(G.1) To examine the ability of STM in capturing different geometric patterns;

(G.2) To examine the posterior estimates of spatial varying transformation parameters under two scenarios, including a no transformation model;

(G.3) To investigate the sensitivity of STM to the specification of $\varphi_k$ and $(-a, b)$;

(G.4) To investigate the sensitivity of STM to the matrix $H_k$;

(G.5) To illustrate the fast convergence of the Gibbs sampler algorithm.

We randomly generated $n = 200$ lattices of size $32 \times 32$ according to model (1), in which we set $\sigma(d) = 0.3$ for all $d$ and $\mathbf{x}_i = (x_{i0}, x_{i1}, x_{i2}, x_{i3})^T$ for $i = 1, \ldots, 200$. The covariates $\mathbf{x}_i$ were generated to mimic real data and include an intercept, a continuous variable, and two categorical variables. They were generated as follows: (i) $x_{i1}$ is generated from $N(5, 1)$; (ii) $x_{i2}$ and $x_{i3}$, respectively, represent the second and third category of a discrete uniform random variable generated from three possible values, each of them representing a category and defined by $x_{iq} = \mathbf{1}(\text{Category } q) - \mathbf{1}(\text{Category } 1)$ for $q = 2, 3$. We generated the values of the transformation parameters $\lambda_d$ from a discrete uniform random variable taking 0.5, 1, or 2. The generated $\Lambda$ structure is presented in the left panel of Figure 1. The parameters in $\boldsymbol{\beta}$ are chosen to have a strong spatial correlation and their images are presented in the panels (a)–(d) of Figure 2.

For the hyperparameters of $\boldsymbol{\beta}$, we chose a noninformative prior for each $\nu_k$ by setting $n_\nu = 10^{-3}$ and $s_\nu^2$. As for the entries of the matrix $H_k$, we set it as in (3) and took the weights as

$\omega_k(d, d') = K(\|d - d'\|_2)$, where $K(u) = \exp(-\frac{1}{2}u^2)\mathbf{1}(u \leq r_0)$ and $r_0 = 2$. For each parameter $\tau_\sigma$, we chose noninformative priors by setting $\delta_0 = 10^{-3}$ and $\gamma_0 = 10^{-3}$. We fixed $\varphi_k$ at 10 that indicates a strong spatial dependency among the components of each $\boldsymbol{\beta}_{(k)}$, and then we set $a = b = 3$ for the hyperparameters of $\lambda_d$.

For each simulated dataset, we ran the Gibbs sampler for 1,000 iterations with 50 burn-in iterations. For the simulated examples, each iteration of the Markov chain takes approximately 2.5 seconds when running on a laptop with an i7 processor, 2.67 GHz, and 8.0 GB of RAM. We summarize some simulation results based on some selected simulation scenarios below, while some additional results obtained from different simulation scenarios are put in the web supplementary materials.

Firstly, Figure 1 reveals that the estimated and true structures of $\mathbf{\Lambda} = \{\lambda_d, d \in \mathscr{D}\}$ show great similarity with each other. As expected, the estimated image $\hat{\mathbf{\Lambda}} = \{\hat{\lambda}_d, d \in \mathscr{D}\}$ is smoother than the true $\mathbf{\Lambda} = \{\lambda_d, d \in \mathscr{D}\}$ since a $U(-3, 3)$ prior is assumed for $\lambda_d$, allowing $\lambda_d$ to be sampled within this interval.

Secondly, we explore whether STM can recover the underlying spatial structure of each coefficient image. See Figure 2 for details. We compare the STM with two other models, including a voxel-wide linear model (panels (e)–(h)) and our STM (1) with $\lambda_d$ fixed at 1 across all voxels (panels (i)–(l)). Figure 2 reveals that the voxel-wide linear model and STM (1) with $\lambda_d$ fixed at 1 cannot capture the pattern of true coefficient images. In contrast, STM (1) substantially improves the estimation of the coefficients, recovering their true geometric patterns, as observed in Figure 2, panels (m)–(p). Moreover, the STM is robust to the choices of the hyperparameters $\varphi_k$ and $(-a, b)$. Furthermore, the correct specification of the matrix $H_k$ can yield good estimates if a reasonable neighborhood system is chosen. Finally, even if the true underlying model does not require spatial transformation parameters, STM can still provide good estimates of $\boldsymbol{\beta}$.

Thirdly, we illustrate the MCMC results for the parameters $\boldsymbol{\beta}$, $\tau_\sigma$ and $\lambda$ at a randomly selected voxel. See Figure 3 for details. The trace plots indicate fast convergence of the Gibbs sampler, confirming its efficiency and good mixing properties. In addition, a more detailed diagnostics analysis is presented in the Web Appendix A. Based on the aforementioned results, we can conclude that the proposed single-site Gibbs sampler algorithm has good mixing properties and reaches convergence rapidly.

## 4. Application to the ADHD dataset

Our model is applied to the Attention Deficit Hyperactivity Disorder data, obtained from the ADHD-200 Consortium, (http://fcon_1000.projects.nitrc.org/indi/adhd200), a self-organized initiative where members from institutions around the world provide de-identified, HIPAA compliant imaging data. The goal of the project is to accelerate the scientific community's understanding of the neural basis of ADHD, which is one of the most common childhood disorders affecting at least 5–10% of school age children and is associated with substantial lifelong impairment. The symptoms include difficulty staying focused and paying attention, difficulty controlling behavior, and hyperactivity (over-activity).

We analyze the imaging data from the New York University (NYU) Child Study Center. There are 219 subjects, 99 controls and 120 diagnosed with ADHD. Among them, 143 are males and 76 are females with an average age of 11.71 and 11.55 years, respectively. We used the high-resolution T1-weighted MRI images that were acquired using the MPRAGE (Magnetization-prepared Rapid Acquisition with Gradient Echo) technique. The original T1-weighted images have size $256 \times 256 \times 198$ mm$^3$ and voxel size of $1.0 \times 1.0 \times 1.0$ mm$^3$.

For each subject, the images were first downsampled to the size of $128 \times 128 \times 99$ mm$^3$. This process reduces the number of voxels while maintaining the image features and properties. Next, the images were processed using HAMMER (Hierarchical Attribute Matching Mechanism for Elastic Registration), a free pipeline developed by the Biomedical Research Imaging Center at UNC (available for downloading at http://www.hammersuite.com). The processing steps include skull and cerebellum removal, followed by tissue segmentation to identify the regions of white matter (WM), gray matter (GM) and cerebrospinal fluid (CSF). Then, registration was performed to warp all subjects to the space of the Jacob template (Kabani et al., 1998; Davatzikos et al., 2001). Finally, a RAVENS map was calculated for each subject. The RAVENS methodology precisely quantifies the volume of tissue in each region of the brain. The process is based on a volume-preserving spatial transformation that ensures that no volumetric information is lost during the process of spatial normalization. In Figure 4, we illustrate the white matter RAVENS images for two randomly selected subjects (panels (a) and (b)). These images were registered to the space of the template shown in panel (c). When we compare subjects in panels (a) and (b) of Figure 4, the image from the subject in panel (b) shows higher brightness inside the green square, reflecting the fact that relatively more white matter is presented in that particular region relative to the template.

We fitted model (1) with the white matter RAVENS images as responses and the covariate vector containing intercept, gender, age (previously standardized) and ADHD diagnostic status (1 for ADHD and −1 for control). Our interest is to identify morphological differences in the brain that are associated with the ADHD outcome, while adjusting for age and gender. As in the simulation study, for the hyperparameters of $\boldsymbol{\beta}$, we chose a noninformative prior for each $\nu_k$ by setting $n_\nu = 10^{-3}$ and $s_\nu^2 = 1$. We fixed $\varphi_k = 10$ and set $\omega_k(d, d') = K(\|d - d'\|_2)$, where $K(u) = \exp(-\frac{1}{2}u^2)1(u \le r_0)$ and $r_0 = 2$. For each parameter $\tau_d$, we chose a noninformative prior by setting $\delta_0 = 10^{-3}$ and $\gamma_0 = 10^{-3}$. For the transformation parameters $\lambda_d$, we set $a = b = 2$. We ran the Gibbs sampler for 1,000 iterations with 50 burn-in iterations. We calculated the posterior mean and a 95% credible interval for the coefficient associated with ADHD outcome at each voxel. To detect important regions of interest, we created a 5% threshold map by mapping whether the 95% credible interval at each voxel contains 0 or not. Finally, we also fitted a no-transformation model, which is the STM with $\lambda_d$ fixed at 1 for all voxels.

An initial exploratory analysis was performed to examine whether the imaging measurements in the RAVENS map follow the Gaussian distribution. Normal probability plots of the intensities from sixteen random voxels are displayed in Figure 5, revealing that for some voxels, the imaging measurements strongly deviate from the Gaussian distribution. Further investigation of the posterior distribution of $\boldsymbol{\Lambda} = \{\lambda_d, d \in \mathscr{D}\}$ reveals that the transformation parameters are different from 1 for nearly 70% of the voxels, based on a 95% credible interval (Figure 6, panel (d)).

We then mapped $\widehat{\boldsymbol{\Lambda}}$ into the template to observe how the transformation parameter varies across the brain. If morphological differences exist in the regions where the transformation parameters are significantly different from 1, then analyzing the imaging data using the standard voxel-wise linear model may lead to spurious conclusions. On the other hand, if the transformation parameters are close to 1 in some regions, the estimates of the STM will be similar to those of the standard voxel-wise linear model in the regions. However, in practice, the location of such regions is unknown.

We compared the results from the STM with those from the no transformation model. Inspecting Figure 7, we are able to detect three large regions of interest, where morphological differences exist, including the right frontal lobe, the left frontal lobe and left parietal lobe. The frontal lobe has been implicated in planning complex cognitive behavior, personality expression, decision making and moderating social behavior (Yang and Raine, 2009) and morphological differences in this region were previously identified in children with ADHD (Sowell et al., 2003). Although the right frontal lobe is noticeable in all panels of Figure 7, the left frontal lobe cannot be seen for the no-transformation model in panel (d) of Figure 7. Thus, without the use of data transformations, we may miss some biologically meaningful regions of interest.

## 5. Discussion

We have proposed a method to model the association between imaging data and clinical outcomes. The proposed model simultaneously overcomes two major limitations of voxel-wise methods that are widely used to model imaging data. First, the lack of normality of imaging measurements is circumvented by proposing a spatial varying Box-Cox transformation model. Second, the voxel-wise methods treat all voxels as independent units, and thus they ignore important spatial smoothness observed in imaging data. We address this issue by assuming a Gaussian Markov random field (GMRF) prior to capture the spatial correlation and spatial smoothness among the regression coefficients in neighboring voxels. We developed an efficient Markov chain Monte Carlo (MCMC) algorithm to sample from the joint posterior distribution of the parameters. Our simulations and real data analysis demonstrate that STM significantly outperforms the standard voxel-wise model in recovering meaningful regions of interest.

## Supplementary Material

Refer to Web version on PubMed Central for supplementary material.

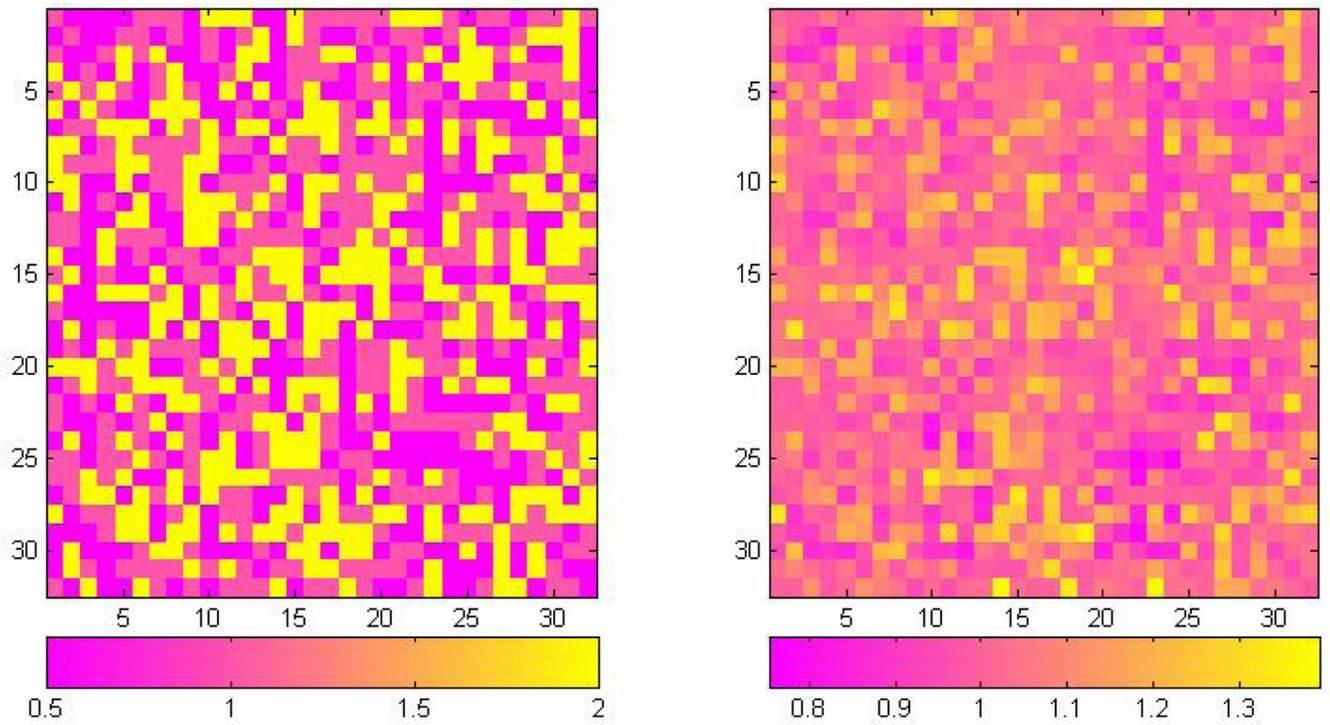

**Figure 1.**
Simulation results: the true $\mathbf{\Lambda} = \{\lambda_d, d \in \mathscr{D}\}$ pattern in the left panel and the estimated pattern in the right panel. Estimated image is smoother compared with the true image due to the nature of the uniform distribution assumed *a priori*.

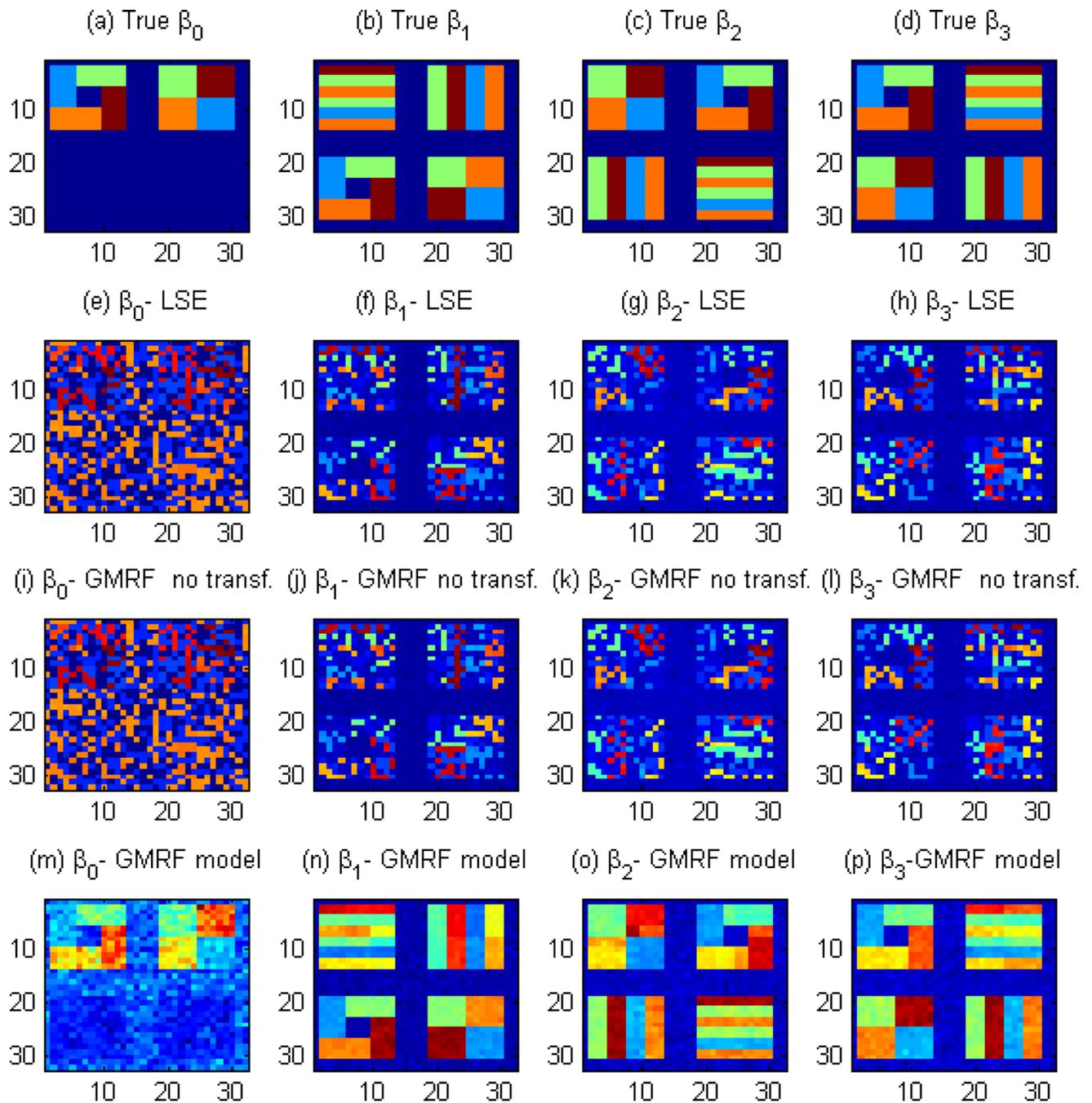

**Figure 2.**
Simulation results on comparison of STM, GMRF with no transformation, and the voxel-wise linear model. Panels (a)–(d) represent the pattern of β used to generate the images; panels (e)–(h) represent the estimated β obtained from the least squares estimator in Matlab; panels (i)–(l) represent the posterior mean of β obtained by fitting a GMRF model with no transformation; and panels (m)–(p) are the posterior mean of β obtained from our STM. The inclusion of the transformation parameter substantially improves the estimation of the true underlying pattern.

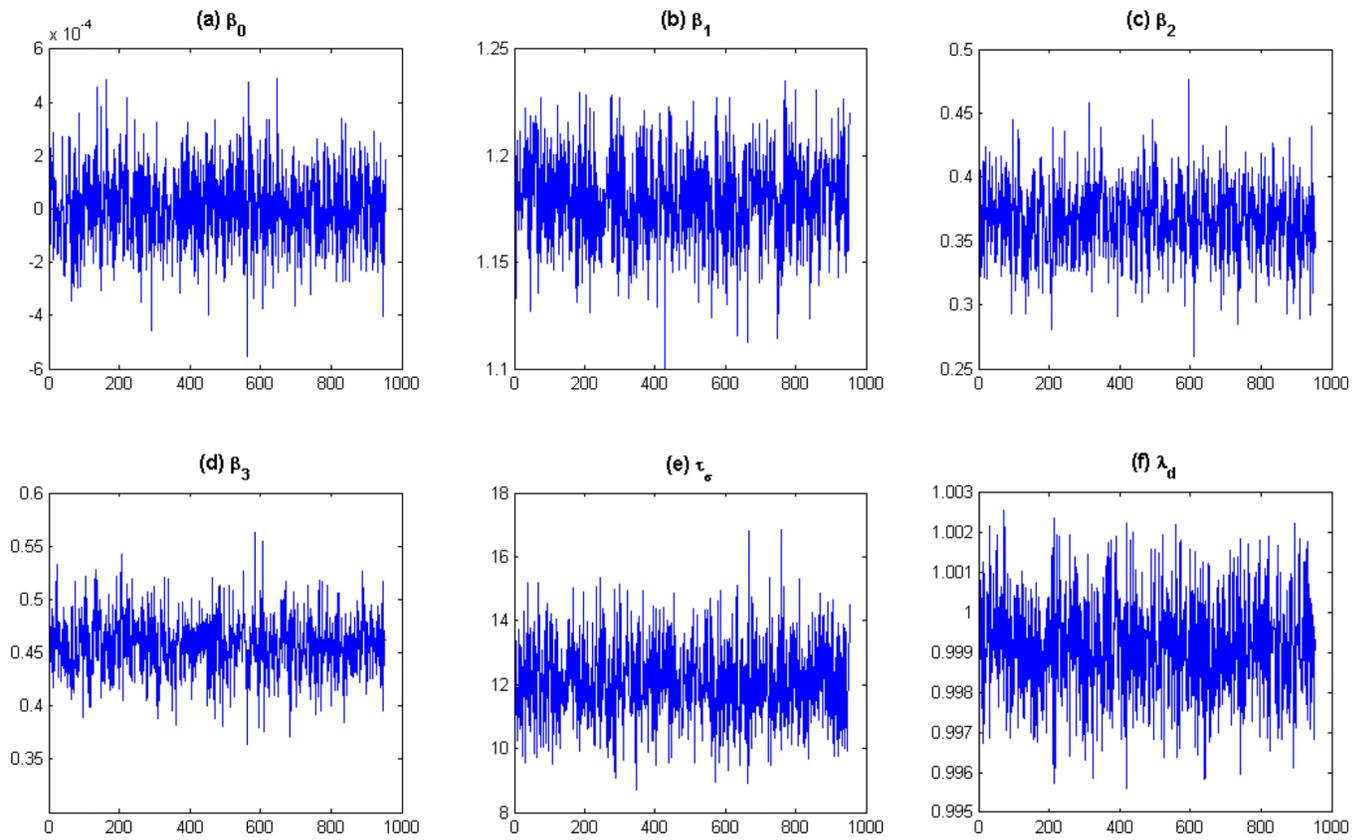

**Figure 3.**
Trace plots for $\boldsymbol{\beta}$, $\tau_\sigma$ and $\lambda$ for a randomly generated voxel. The results are for a 1000 iterations of the MCMC algorithm and a burn-in sample of 50. The trace plots indicate a fast convergence of the algorithm, confirming its efficiency and good mixing properties.

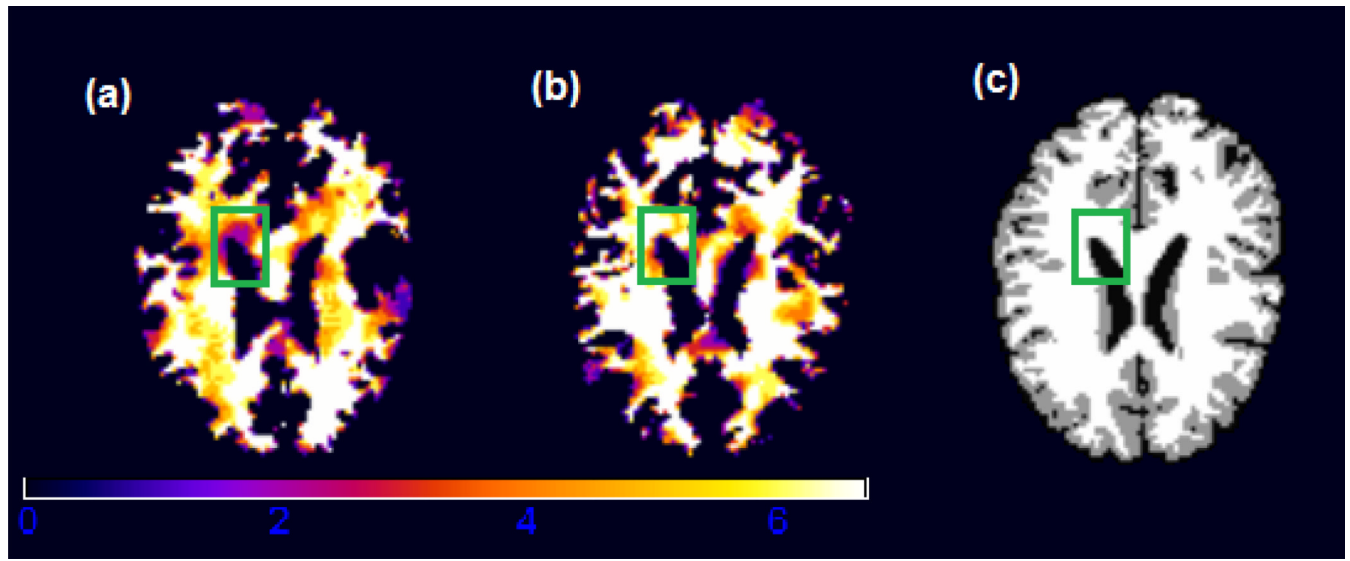

**Figure 4.**
White matter RAVENS map for two randomly selected children from the ADHD study. The image from subject b shows a higher brightness inside the green square, reflecting the fact that for the brain of subject b, relatively more white matter was forced to fit the same template (panel (c)) at that particular region.

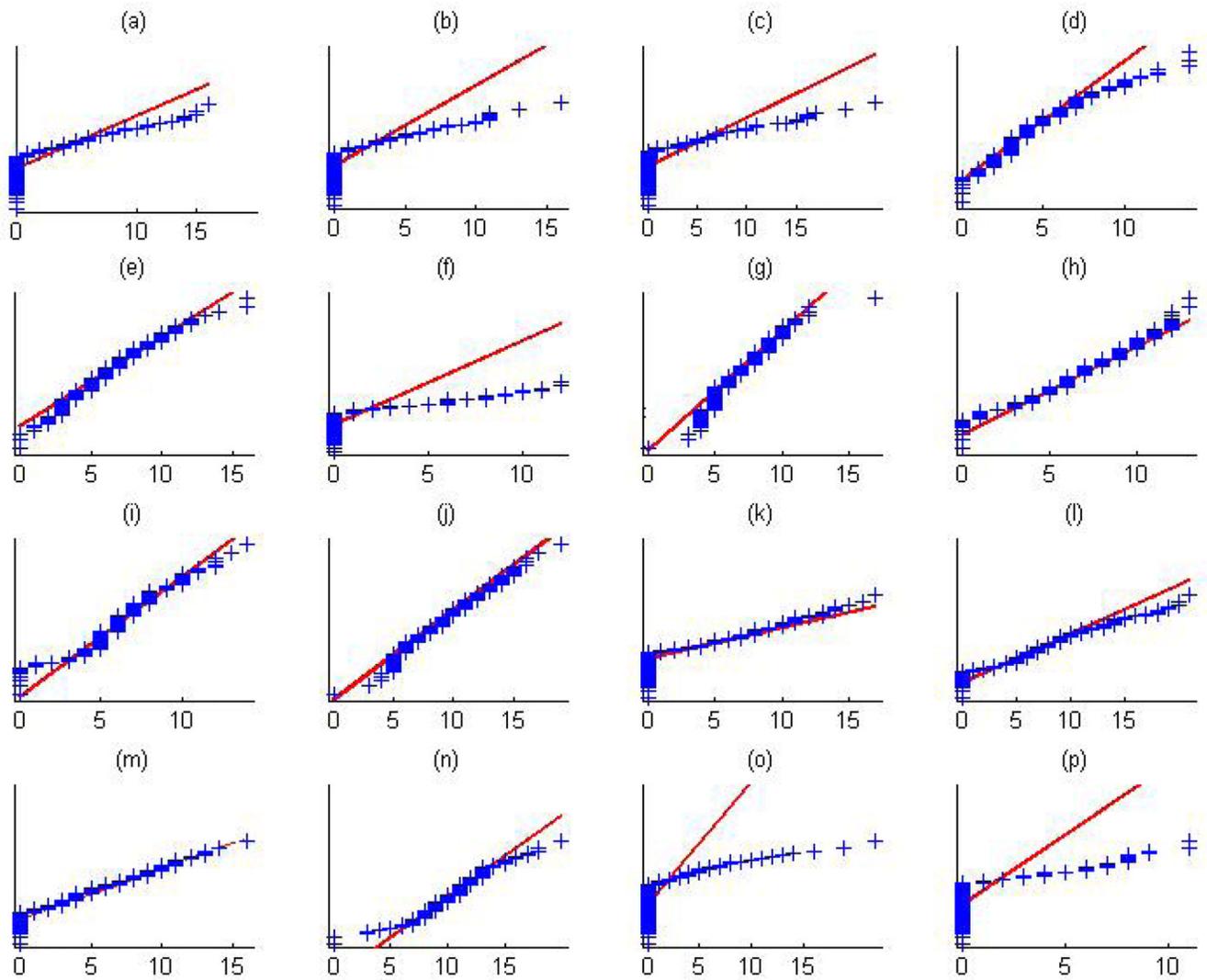

**Figure 5.**
ADHD data analysis results: normal probability plots of sixteen random voxels revealing that the imaging measurements extracted from the RAVENS map deviate from the Gaussian distribution.

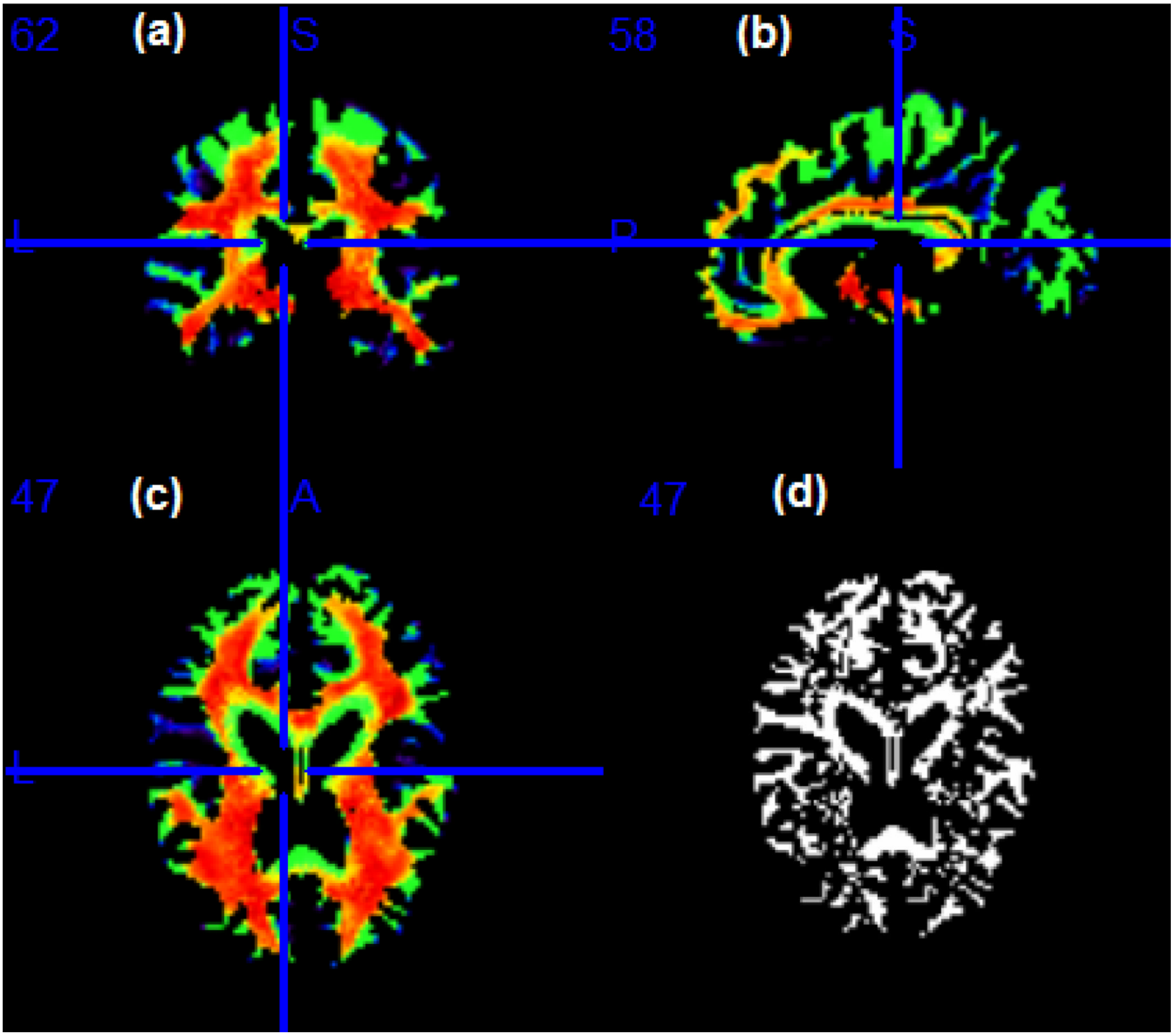

**Figure 6.**
ADHD data analysis results: selected slices showing the estimated $\hat{\boldsymbol{\Lambda}}$ for the imaging data obtained from the white matter RAVENS map. Panels (a)–(c) represent respectively, a coronal, sagittal and axial view of selected slices of the brain. The blue line indicates where the coronal and sagittal slices meet the plane in (c); panel (d) shows the same axial slice as in (c) and represents the location in the brain where $\boldsymbol{\Lambda} = \{\lambda_d, d \in \mathscr{D}$ are different from 1, based on a 95% credible interval.

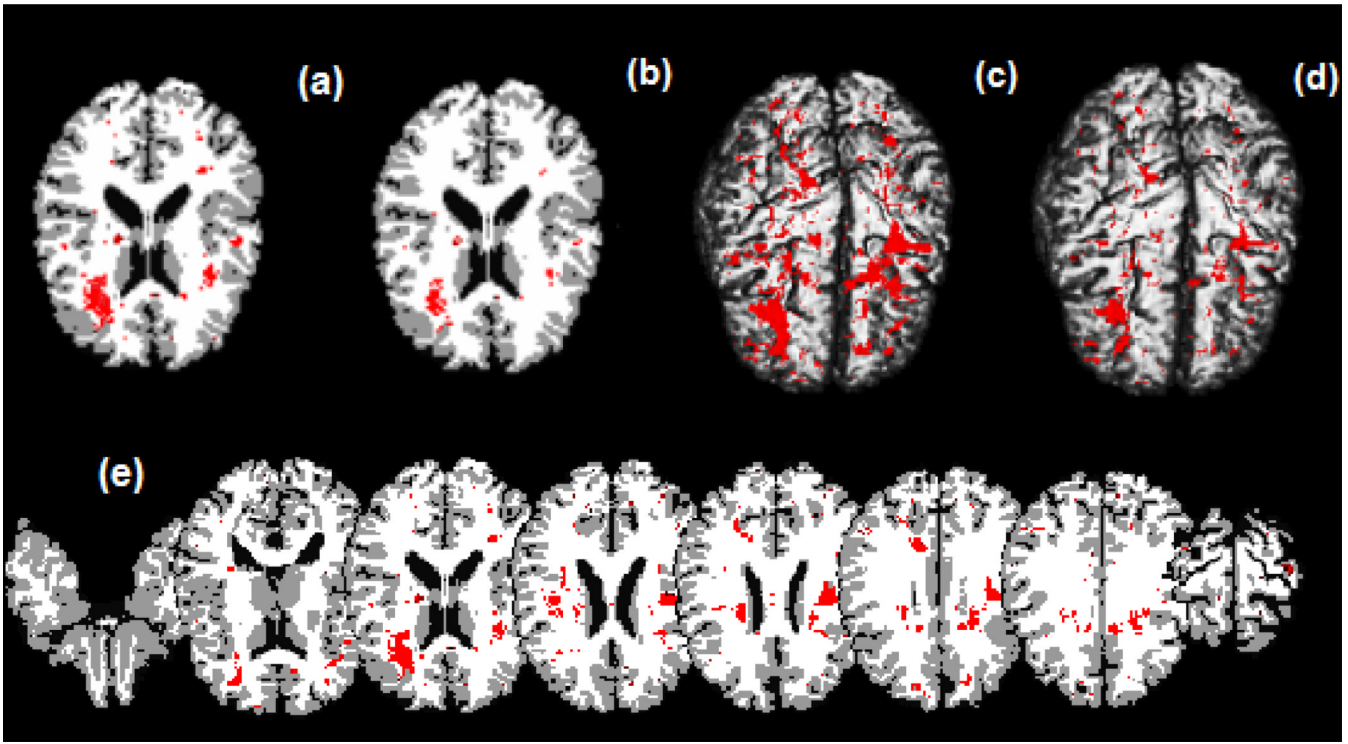

**Figure 7.**
ADHD data analysis results. **Top panels**: significant regions in the brain where there exists a morphological difference between children with ADHD and children who do not have the disorder, based on a 95% credible interval. Panel (a) is a selected axial slice of the STM estimate overlaid on the Jacob template; (b) is the same selected slice showing the estimates of the spatial model with the transformation parameters Λ fixed and equal to 1 for all voxels also overlaid on the template; (c) and (d) are, respectively, the results of a 3D rendering of the STM and of the no transformation model both overlaid on the Jacob template. **Bottom panel**: (e) shows selected axial slices of the STM estimates overlaid on the template. Red areas show the significant regions in the brain where there exists a morphological difference between children with ADHD and children who do not have the disorder.